\documentclass[aps,prd,twocolumn, nofootinbib,tightenlines]{revtex4-1}

\usepackage{bm, amsmath,amssymb,amsfonts,slashed,array,multirow,graphicx}
\usepackage[vcentermath,noautoscale]{youngtab}
\Yboxdim{8pt}

\newcommand{\fund}{\yng(1)}
\newcommand{\afund}{\overline{\yng(1)}}

\newcommand{\abfund}{\overline{\yng(1,1)}}
\newcommand{\twosym}{\yng(2)}

\newcommand{\Gc}{G_{\rm{c}}}
\newcommand{\GF}{G_{\rm{F}}}
\newcommand{\GSM}{G_{\rm{SM}}}
\newcommand{\GFp}{G_{\rm{F}}^\prime}
\newcommand{\UF}{\rm{U}(1)_{\rm{F}}}
\newcommand{\U}[1]{\rm{U(1)}}

\newcommand{\Mpl}{M_{\rm pl}}
\newcommand{\gs}{g_{*_{\rm S}}}
\newcommand{\id}{\bm{1}}

\newcommand{\Nl}{N^h}
\newcommand{\fdm}{f_{\rm dm}}

\usepackage{hyperref}	
\begin{document}
\title{A Dynamical Framework for KeV Dirac Neutrino Warm Dark Matter}
\author{Dean J. Robinson}
\email{djrobinson@berkeley.edu}
\affiliation{Department of Physics, University of California, Berkeley, CA 94720, USA}
\affiliation{Ernest Orlando Lawrence Berkeley National Laboratory,
University of California, Berkeley, CA 94720, USA}
\author{Yuhsin Tsai}
\email{yhtsai@ucdavis.edu}
\affiliation{Department of Physics, University of California, Davis, CA 95616, USA}

\begin{abstract}
If the source of the reported $3.5$~keV x-ray line is a sterile neutrino, comprising an $\mathcal{O}(1)$ fraction of the dark matter (DM), then it exhibits the property that its mass times mixing angle is $\sim \mbox{few} \times 10^{-2}$~eV, a plausible mass scale for the active neutrinos. This property is a common feature of Dirac neutrino mixing. We present a framework that dynamically produces light active and keV sterile Dirac neutrinos, with appropriate mixing angles to be the x-ray line source. The central idea is that the right-handed active neutrino is a composite state, while elementary sterile neutrinos gain keV masses similarly to the quarks in extended Technicolor. The entire framework is fixed by just two dynamical scales and may automatically exhibit a warm dark matter (WDM) production mechanism -- dilution of thermal relics from late decays of a heavy composite neutrino -- such that the keV neutrinos may comprise an $\mathcal{O}(1)$ fraction of the DM. In this framework, the WDM is typically quite cool and within structure formation bounds, with temperature $\sim \mbox{few}\times 10^{-2}~T_\nu$ and free-streaming length $\sim$ few kpc. A toy model that exhibits the central features of the framework is also presented.
\end{abstract}
\maketitle

\section{Introduction} 
Recently two groups \cite{Bulbul:2014sua,Boyarsky:2014jta} reported the detection of a $3.55$~keV x-ray line in galaxy clusters and the Andromeda galaxy. Among various recently discussed scenarios \cite{Allahverdi:2014dqa,Babu:2014pxa,Cicoli:2014bfa,Demidov:2014hka,Higaki:2014zua,Kolda:2014ppa,Kong:2014gea,Krall:2014dba,Lee:2014xua,Nakayama:2014cza,Nakayama:2014ova,Queiroz:2014yna,Abazajian:2014gza,Baek:2014qwa,Ishida:2014dlp,Modak:2014vva,Cline:2014eaa,Okada:2014zea,Lee:2014koa,Dudas:2014ixa,Ko:2014xda,Tsuyuki:2014aia,Baek:2014poa,Chakraborty:2014tma,Chen:2014vna,Chiang:2014xra,Conlon:2014wna,Conlon:2014xsa,Dubrovich:2014xaa,Dutta:2014saa,Ishida:2014fra,Nakayama:2014rra,Riemer-Sorensen:2014yda}, this signal may be the signature of a $m_d \simeq 7.1$~keV sterile neutrino that mixes with the active neutrinos. Such sterile neutrinos are a well-studied candidate for warm dark matter (WDM) (see e.g. Refs.~\cite{Boyarsky:2009ix,Kusenko:2009up} for a review). Whether WDM better accommodates various small-scale structure formation controversies compared to the $\Lambda$CDM framework is itself controversial (see e.g. \cite{Bode:2000gq,Barkana:2001gr, Somerville:2003sh,Colin:2007bk,Boyanovsky:2008co,Zavala:2009ms,deVega:2010yk, deVega:2013ysa,Weinberg:2013aya,Markovic:2013iza, Viel:2013fqw, Schultz:2014eia}), but, at least for the moment, it remains a competitive DM paradigm for sufficiently large thermal masses \cite{Schultz:2014eia}. 

In almost all keV neutrino models, the neutrino masses are generated by a seesaw, with the consequence that the sterile neutrinos are Majorana (for a review of various approaches to keV neutrino model-building, including Froggatt-Nielsen, neutrino flavor and other modified seesaw mechanisms, see Ref.~\cite{Merle:2013gea} and references therein). With this in mind, we point out here that this new x-ray line has an interesting feature: If the x-ray source is a sterile neutrino comprising 100\% of the dark matter (DM), then the observed flux of the x-ray line corresponds to an active-sterile mixing angle $\sin^2(2\theta_d) \simeq 6.8\times10^{-11}$ \cite{Bulbul:2014sua}. One then observes that $m_d\theta_d \simeq 0.03$~eV, a plausible mass scale for the active neutrinos, $m_a$. That is, the mixing angle $\theta_d \simeq m_a/m_d$. While hard to achieve in Majorana neutrino models, this property is a common feature of mixing between \emph{Dirac} neutrinos. This observation strikingly suggests that the x-ray source could be a sterile Dirac neutrino.

In this Note, we present a framework that dynamically produces sub-eV active Dirac neutrinos and keV sterile neutrinos, with a mixing angle automatically at the observed scale. With special choices, the framework can automatically include a late-decaying heavy neutrino. The lifetime and mass scales of this heavy neutrino, along with the corresponding entropy production, occur in just the right amounts such that: a thermal relic of the keV sterile neutrinos is diluted to an $\mathcal{O}(1)$ fraction or all of the observed DM relic abundance; the diluted relic has a temperature suitable to be cool WDM, within current structure formation bounds; and the Standard Model (SM) plasma is reheated above the big-bang nucleosynthesis (BBN) temperature scale. Up to $\mathcal{O}(1)$ numbers, the entire framework is fixed by just two dynamical scales -- one is a UV completion scale $M$ ($\sim 5\times10^4$~TeV), the other is a confinement scale, $\Lambda$ ($\sim10$~TeV). 

This framework was first presented in Ref.~\cite{Robinson:2012wu}. The central idea is that the right-handed active neutrinos are composite states of a confining hidden sector \cite{ArkaniHamedGrossman:1999,Okui:2004xn,GrossmanTsai:2008,Grossman:2010iq,McDonald:2010jm,Duerr:2011ks}. The sterile keV neutrinos are, in contrast, elementary states, that gain their masses in a similar manner to the quarks in extended Technicolor theories. In this Note, we develop this framework further for a specific case of the general class of theories considered in Ref.~\cite{Robinson:2012wu}. In particular we present a different, simpler DM production mechanism for this framework -- alternatives were presented in \cite{Robinson:2012wu}. A toy model that exhibits the main features of this framework is also presented.

\section{Framework} 
\subsection{Ingredients}
Details on the theoretical foundations of this framework can be found in Refs~\cite{Robinson:2012wu,GrossmanTsai:2008}; in the following we provide a condensed synopsis. 

Along with the SM field content and gauge groups $\GSM$, suppose that there exists a hidden sector that is charged under a hidden confining group $\Gc$ and a hidden flavor symmetry $\GF$. We denote the SM content by $\sigma$, and imagine that the hidden sector is comprised of chiral fermions denoted by $\chi$ and $\xi$ such that
\begin{equation}
		\sigma \sim \GSM\otimes\GF~, \qquad \chi \sim \Gc\otimes\GF~,\qquad \xi \sim \GF~.
\end{equation}
We consider this to be a low-energy effective field theory below some UV completion scale, $M$, so that the SM and hidden sectors interact via $M$-scale irrelevant operators.

At a scale $\Lambda \ll M$, $\Gc$ becomes strongly coupled, inducing a confining phase transition (CPT) of the $\chi$ into bound states. We therefore call $\chi$ preons, in concordance with the usual language. Confinement of the preons typically breaks the hidden flavor group $\GF \to \GFp$. If $\GFp$ is non-trivial and the $\chi$ sector has non-trivial $\GFp$ anomalies -- cancelled by the non-confining sectors -- then anomaly matching between the free and confined phases implies the existence of chiral bound states \cite{tHooft:1980,DimopoulosRabySusskind:1980,RabyDimopoulosSusskind:1980tg}. The $\xi$ are spectators to the confinement, that are necessarily included to cancel the $\GF$ anomalies. 

With appropriate choices for the hidden sector structure, one can ensure simultaneously: (i) $\GFp = \UF$;  (ii) there are precisely three chiral bound states, $n_R$, all with the same $F$ charge; (iii) $F$ is a linear combination of $B-L$ and hypercharge; (iv) the Higgs carries $F$ charge, such that $\mbox{U(1)}_{\rm EM}\otimes\mbox{U(1)}_{B-L}$ remain after electroweak symmetry breaking (EWSB); (v) the $n_R$ have appropriate $F$ charge to be right-handed neutrinos, i.e. they form the $L^c_LH^\dagger n_R$ yukawas. Generically, along with the three $n_R$, there may be fermionic bound states (spectators) $N_L^c$ and $N_R$ ($\xi_L^c$ and $\xi_R$) which furnish Dirac multiplets, and have the correct $F$ charge to be sterile neutrinos. Taken together, one obtains a spectrum of chiral and Dirac states, such as the one shown in Table \ref{tab:CBS}. One may check that a $\UF$ with such charges is non-anomalous, and in fact $2Y - F = B-L$. After EWSB by the Higgs field, $H$, $\mbox{U(1)}_{\rm B-L}$ remains a low energy symmetry of the framework. 

\def\arraystretch{1.3}
\begin{table}[t]
\begin{center}
\begin{tabular*}{0.95\linewidth}{@{\extracolsep{\fill}}|c|ccccccc|cccc|}
	\hline
	 $\quad$ & $H$ & $L^c_L$ & $E_R$ & $Q_L^c$ & $U_R$ & $D_R$ & $n_R$ & $N_L^c$ & $N_R$ & $\xi_L^c$ & $\xi_R$\\
	\hline\hline
	F & $+1$ & 0 & $-1$ & 0 & $1$ & $-1$ & $+1$ & $-1$ &  $+1$ & $-1$ & $+1$ \\
	Y & $1/2$ & $1/2$ & $-1$ & $-1/6$ & $2/3$ & $-1/3$ & $0$ & $0$ & $0$  & $0$ & $0$\\
	\hline
\end{tabular*}
\end{center}
\caption{$\UF$ and hypercharge assignments to the heavy sterile neutrinos, sterile neutrino spectators and SM content. The $n_R$, $N_{L,R}$ and $\xi_{L,R}$ are SM sterile by construction. The $N_{L,R}$ and $\xi_{L,R}$ form Dirac fermions, while the remainder of the fermions shown here are chiral before EWSB.}
\label{tab:CBS}
\end{table}

\subsection{Mass and Mixing Angles} 
In order to avoid flavor constraints, necessarily $M \gg v$. It is therefore convenient to define two parameters
\begin{equation}
	\label{eqn:PD}
	\epsilon \equiv \Lambda/M \ll1~,\qquad \vartheta \equiv v/M \ll 1~.
\end{equation}
We assume the fermionic bound states that form sterile neutrinos contain three preons and the symmetry breaking condensate has two preons: in the notation of Ref.~\cite{Robinson:2012wu} this is a `$(3,2)$' model. 

After confinement, Dirac mass terms between the active lepton doublet and sterile neutrinos are generated from the operators
\begin{align}
L^c_LH^\dagger \chi^3/M^3 & \to \frac{\epsilon^3}{(4\pi)^2} L^c_LH^\dagger n_R \quad \mbox{or} \quad  \frac{\epsilon^3}{(4\pi)^2} L^c_LH^\dagger N_R \notag\\
L^c_L H^\dagger \chi^2\xi/M^3 & \to  \frac{\epsilon^3}{(4\pi)^2} L^c_LH^\dagger \xi_R~.
\end{align}
Here the factors of $4\pi$ arise from na\"\i ve dimensional analysis matching between the free and confined phases \cite{Cohen:1997rt,Manohar1984189}. (The factors of $4\pi$ were omitted in Ref. \cite{Robinson:2012wu}.) The spectators obtain Dirac masses through irrelevant couplings to the condensate vev, i.e. 
\begin{equation}
	\label{eqn:LMT}
	\xi \langle \chi^2 \rangle \xi/M^2  \to  \frac{\Lambda\epsilon^2}{(4\pi)^2} \xi^c_L\xi_R~.
\end{equation}
There is no generic name in the literature for spectators that gain masses in this manner, not to mention ones that further act as sterile neutrinos. We propose to call these spectators `laurinos'\footnote{Latin-Italian hybrid for a sprig of bay laurel, and by extension, a `small victory' (cf. \emph{laureola}).}, because they enhance the flavor structure of the hidden sector, and end up being the keV WDM candidates.

Motivated by the large $N_c$-counting rules for QCD baryons \cite{Manohar:1998xv}, we expect the three-preon massive bound states $N_{L,R}$ to typically have a Dirac mass $\sim 3\Lambda$. We assume that the $\GF$ structure of the laurinos does not admit Dirac mass cross-terms with the bound states, and moreover that the proton decay operator $uude/M$ is forbidden by details of the UV theory above $M$.

After EWSB, these operators lead to a mass term of block-matrix form \cite{Robinson:2012wu}
\begin{equation}
	\label{eqn:BMMT}
	\frac{\Lambda}{(4\pi)^2} \begin{pmatrix}	\nu^c_L & \xi^c_L & N^c_L \end{pmatrix} \begin{pmatrix} \vartheta \epsilon^2 & \vartheta \epsilon^2 & \vartheta \epsilon^2\\ 0 &\epsilon^2  & 0 \\ 0 & 0 & 3(4\pi)^2\end{pmatrix}\begin{pmatrix}n_R \\ \xi_R \\ N_R\end{pmatrix}~.
\end{equation}
We denote the mass eigenstates by $\nu^{a,d,h}_{L,R}$, the superscripts anticipating the names active, dark and heavy respectively. The corresponding spectrum is
\begin{equation}
	\label{eqn:MS}
	m_a \sim \frac{v\epsilon^3}{16\pi^2} \qquad m_d \sim \frac{\Lambda \epsilon^2}{16\pi^2} \qquad m_h \sim 3\Lambda~,
\end{equation}
while the $\nu^a$ -- $\nu^d$ mixing angle $\theta_{d} \sim \vartheta$. In more detail, at leading order in $\epsilon$ and $\vartheta$ the mass eigenstates are
\begin{align}
	\begin{pmatrix} \nu^a_L \\ \nu^d_L \\ \nu^h_L \end{pmatrix} 
	& \sim \begin{pmatrix} 1 & \vartheta & \frac{\vartheta\epsilon^2}{48\pi^2} \\ \vartheta  & 1 & \simeq0 \\ \frac{\vartheta\epsilon^2}{48\pi^2} & \simeq0 & 1 \end{pmatrix}\begin{pmatrix} \nu_L \\ \xi_L \\ N_L \end{pmatrix}~,\notag\\
	\begin{pmatrix} \nu^a_R \\ \nu^d_R \\ \nu^h_R \end{pmatrix} 
	& \sim \begin{pmatrix} 1 & \vartheta^2 & \simeq0 \\ \vartheta^2  & 1 & \simeq0 \\  \simeq0 & \simeq0 & 1 \end{pmatrix}\begin{pmatrix} n_R \\ \xi_R \\ N_R \end{pmatrix}~. \label{eqn:ME}
\end{align}
Fixing $\epsilon \sim 3\times10^{-4}$, $\Lambda \sim 15$~TeV, we immediately produce $m_a \sim 0.03$~eV, $m_d \sim 5$~keV and $\sin^2(2\theta_{d}) \sim 5\times10^{-11}$ up to $\mathcal{O}(1)$ coefficients. These scales closely match the observed x-ray line and active neutrino mass-squared splitting results.

\section{A Toy Model} 
For the sake of concreteness, we write in Table \ref{tab:SMF}  a toy model that reproduces the salient features of this framework. Here $\Gc =  \rm{SU}(6)_1\otimes\rm{SU}(6)_2$ and $\GF = \rm{SU}(2)_{\rm{F}}\otimes\UF$. The preonic sector consists of two subsectors, charged under different confining $\rm{SU}(6)$ gauge interactions, that interact only via $M$-scale interactions. We assume that both $\rm{SU}(6)$ gauge interactions confine at approximately the same scale $\Lambda$, noting that the $\beta$ functions for each gauge coupling differ by only a few percent. One preonic subsector has non-trivial $\rm{SU}(2)_{\rm{F}}\otimes\UF$ anomalies such that three SM flavors are required for anomaly cancellation, along with laurinos, too. 

\def\arraystretch{1.3}
\begin{table}[t]
\newcolumntype{C}{ >{\centering\arraybackslash $} m{0.8cm} <{$}}
\newcolumntype{D}{ >{\raggedright\arraybackslash $} m{1.2cm} <{$}}
\begin{tabular*}{0.95\linewidth}{@{\extracolsep{\fill}}|C|cC|CCCC|D|}
\hline
 & \# & \mbox{Field} & \rm{SU}(6)_1 &\rm{SU}(6)_2 & \rm{SU}(2)_{\rm{F}} & \UF & \GSM\\
\hline\hline
\multirow{4}{*}{\rotatebox{90}{\parbox{2cm}{Preons}}} 
& 2 & \chi_0& \id & \fund & \id & +2/3 & -\\
& 2 & \chi_1& \fund & \id & \id & -5/6 & -\\
& 1 & \chi_2 & \id & \abfund & \id & -1/3 & -\\
& 1 & \chi_3 & \afund & \id & \fund & +5/6 & -\\[5pt]
\hline
\multirow{6}{*}{\rotatebox{90}{\parbox{0.5cm}{SM}}} 
& 3 & E_R & \id &  \id & \id & -1 & \big(\id,\id\big)_{-1}\\
& 3 & L^{c}_L &  \id & \id & \id & 0 & \big(\id,\bm{2} \big)_{+\frac{1}{2}}  \\
& 3 & U_R & \id &  \id & \id & +1 & \big(\bm{3},\id\big)_{+\frac{2}{3}} \\
& 3 & D_R &  \id & \id & \id & -1  & \big(\bm{3},\id\big)_{-\frac{1}{3}}\\
& 3 & Q^{c}_L & \id &  \id & \id & 0 &\big(\bar{\bm{3}},\bm{2}\big)_{-\frac{1}{6}}\\
& 1 & H &  \id & \id & \id & +1&  \big(\id,\bm{2}\big)_{+\frac{1}{2}} \\[5pt]
\hline
\multirow{2}{*}{\rotatebox{90}{\parbox{1cm}{DM}}} 
& 3 & \xi_0 &  \id & \id & \fund & +1 & -\\
& 2 & \xi_1 & \id &  \id & \twosym & -1 & -\\
\hline
\end{tabular*}
\caption{A toy model of the composite neutrino framework in the unconfined phase. The field $H$ is the EWSB scalar; all other fields are fermionic. $\GSM = \mbox{SU(3)}_{\rm c}\otimes\mbox{SU(2)}_{\rm L}\otimes\mbox{U(1)}_{\rm Y}$. Note that choosing $\rm{SU}(6)_1 = \rm{SU}(6)_2$, so that $\Gc$ is simple, also leads to a consistent, non-anomalous theory.}
\label{tab:SMF}
\end{table}

\def\arraystretch{1.3}
\begin{table}[t]
\newcolumntype{C}{ >{\centering\small\arraybackslash $ } m{1.3cm}  <{$}}
\begin{tabular*}{0.85\linewidth}{@{\extracolsep{\fill}}|C|cC|CC|}
\hline
& \# &  \mbox{Content} &  \rm{SU}(2)_{\rm{F}} & \UF \\
\hline\hline
\multirow{3}{*}{Fermions}
& 4 & \chi_0^2\chi_2 & \id & +1\\
& 1 & \chi_2^3 & \id & -1\\
& \ldots &\ldots&\ldots&\ldots\\
\hline
\multirow{2}{*}{Scalars} & 2 & \chi_1\chi_3 & \fund & 0\\
&\ldots &\ldots&\ldots&\ldots\\
\hline
\end{tabular*}
\caption{Spectrum of scalar and right-handed fermionic bound states arising from the preonic sector in Table \ref{tab:SMF}. Other fermionic bound states furnish Dirac or Majorana multiplets after the CPT.}
\label{tab:SMFC}
\end{table}

The corresponding bound state spectrum for this theory is presented in Table \ref{tab:SMFC}. We assume $\langle \chi_1\chi_3\rangle \not= 0$, so that $\rm{SU}(2)_{\rm{F}}$ is broken completely, producing the symmetry breaking pattern
\begin{equation}
	 \rm{SU}(2)_{\rm{F}}\otimes\UF \to \UF~.
\end{equation}
To analyze the residual bound state chiral structure, this theory can be treated as a tumbling gauge theory \cite{DimopoulosRabySusskind:1980,RabyDimopoulosSusskind:1980tg}. In this picture, three of the four $\chi_0^2\chi_2$ bound states are heuristically identified as the chiral bound states. One may check that they satisfy the $\UF$ anomaly matching criterion. 

Now, the four right-handed fermionic bound states $\chi_0^2\chi_2$ have $\UF$ charge $+1$ -- the correct charges to be three $n_R$ and one $N_R$ -- while the single $\chi_2^3$ has charge $-1$ and can be $N_L^c$.  Other three-preon bound states form massive Dirac or even Majorana fermions, too, but they cannot act as neutrinos, as they have a different $\UF$ charge.  If we limit the bound states to only three preons each, then there is just one heavy sterile neutrino in this toy model, along with the three chiral $n_R$. Note that the theory also includes six elementary $\xi_{L,R}$ with correct charges to act as sterile Dirac neutrinos, too.

\begin{figure}[t]
	\includegraphics[width= 4.5cm]{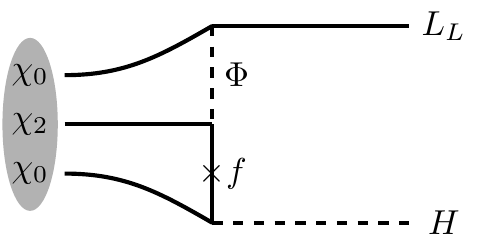}
	\caption{$N_R$ or $n_R$ coupling to $L_L$ and $H$. The scalar mediator $\Phi$ and Dirac mediator $f_{R,L}$ have masses $\sim M$ and charges $(\id,\bar{\bm{6}};\id,-2/3;\id,\bm{2},-1/2)$ and $(\id,\bar{\bm{6}};\id,1/3;\id,\bm{2},1/2)$ with respect to $\Gc\otimes\GF\otimes\GSM$.}
	\label{fig:NLH}
\end{figure}

As desired, the mass terms $L_L^c H^\dagger (\chi^2_0\chi_2)/M^3$ and $L_L^cH^\dagger (\chi_1\chi_3) \xi_0/M^3$ are permitted by the flavor symmetries. As an example, the former can be generated via electroweak-charged $M$-scale mediators, with tree-level contribution shown in Fig. \ref{fig:NLH}. Moreover, we may have laurino mass terms through irrelevant coupling to the condensate, i.e. $\xi_0\langle\chi_1\chi_3\rangle \xi_1/M^2$. However, the cross-terms $(\chi_2^3)\xi_0$  and $(\chi_0^2\chi_2) \xi_1$ are forbidden by the SU(2)$_{\rm F}$ flavor symmetry.

The Dirac fermion $N_{L,R}$ formed between $\chi_0^2\chi_2$ and $\chi_2^3$ couples only through $M$-scale interactions with the three massless hidden pions, associated with the spontaneous breaking of SU(2)$_{\rm F}$ by $\langle \chi_1\chi_3 \rangle$: it is for this reason we chose a toy model with a semi-simple $\Gc$. Moreover, the $\GF$ structure of the toy model ensures the composite neutrino-pion couplings only occur through higher-order operators in chiral perturbation theory -- $N_L^cN_R$ is a $\rm{SU}(2)_{\rm F}$ singlet, but $\chi_1\chi_3$ is not. This suggests that these operators may be suppressed by large powers of $\epsilon$, though the actual degree of suppression depends on the UV theory above $M$. In this discussion we assume composite neutrino-pion interactions can be neglected.

\section{Thermal History}
\subsection{Parameters}
Let us now turn to consider the thermal history of this framework, and in particular whether the laurinos can be sterile neutrino WDM. In this framework, all physical quantities are determined up to $\mathcal{O}(1)$ numbers by $\Lambda$ and $\epsilon$, together with the electroweak and Planck scales, $v$ and $\Mpl$ respectively. 

Let the observed x-ray line source be a single laurino species that comprises a fraction $\fdm$ of the DM in the source object, and has mixing angle $\theta_f \ll 1$. The observed x-ray flux, in terms of the source object DM column density $S_{\rm dm}$ and laurino decay rate $\Gamma_d$,
\begin{equation}
	\dot{N} \sim \fdm S_{\rm dm} \frac{\Gamma_d}{m_d} \propto \fdm \frac{M_{\rm dm}}{ d_{\rm dm}^2} m_d^4 \theta_f^2~,
\end{equation}
so that the mixing angle
\begin{equation}
	\theta_f = \theta_d/\sqrt{\fdm}~,
\end{equation}
where, as above, $\theta_d~(\simeq 4 \times 10^{-6})$ is the corresponding mixing angle assuming the x-ray line source is 100\% of the DM, and that the source object DM mass $M_{\rm dm}$ and luminosity distance $d_{\rm dm}$ are correctly estimated.

The parameters $\Lambda$ and $\epsilon$ are now uniquely determined by the observed $m_d$ and $\theta_d$, via \eqref{eqn:PD} and \eqref{eqn:MS}
\begin{equation}
	\label{eqn:LEDR}
	 m_d = \frac{\alpha_m}{(4\pi)^2} \Lambda \epsilon^2~,\qquad \mbox{and} \qquad \theta_d =  \alpha_\theta \sqrt{\fdm} v \epsilon/ \Lambda~, 
\end{equation}
where hereafter $\alpha_i$ denote $\mathcal{O}(1)$ numbers, and we define $\alpha_{\theta f} \equiv \alpha_\theta \sqrt{\fdm}$, as these always appear together. We shall keep track of these numbers for the sake of generality: In what follows we denote the functional form of physical quantities, up to $\mathcal{O}(1)$ numbers, with a `$\sim$'. The corresponding central numerical value with restored $\alpha_i$ dependence is denoted by a `$\to$'. 

The solutions to the two relations \eqref{eqn:LEDR} are
\begin{align}
	\Lambda & \sim \bigg(\frac{v^2m_d}{\theta_d^2}\bigg)^{1/3} \to  13\bigg(\frac{\alpha^2_{\theta f}}{\alpha_m}\bigg)^{1/3}\mbox{TeV}~,\notag\\
	\epsilon & \sim \bigg( \frac{\theta_d m_d}{v}\bigg)^{1/3} \to \frac{3\times10^{-4}}{ (\alpha_m\alpha_{\theta f})^{1/3}}~.\label{eqn:LE}
\end{align}
From eq. \eqref{eqn:MS}, writing the active neutrino masses as $m_{\nu} = \alpha_\nu v \epsilon^3/(4\pi)^2$, then it immediately follows that
\begin{equation}
	m_\nu \sim m_d\theta_d \to 0.028(\alpha_\nu/\alpha_m\alpha_{\theta f})~\mbox{eV}~.
\end{equation}
We emphasize that the scale of the active neutrino masses is a prediction of this framework, for $m_d$ and $\theta_d$ at their observed values. It also proves convenient to define
\begin{equation}
	\label{eqn:DD}
	\delta \equiv \frac{2\pi^{3/2}}{3\sqrt{5}} \frac{\Lambda}{\epsilon^4 \Mpl} \sim \frac{v^2}{m_d\theta_d^2\Mpl} \to 0.22\alpha_m\alpha_{\theta f}^2~,
\end{equation}
which simplifies several expressions below.

\subsection{Decoupling} 
For $\Lambda$ and $\epsilon$ at these scales, freeze out of $M$-scale $2 \to 2$ interactions determines the decoupling between the SM, preonic, and laurino sectors \cite{Robinson:2012wu}. The corresponding decoupling temperature
\begin{equation}
	\label{eqn:TDC}
	T_{\rm dec} \sim  (\gs^{\rm dec})^{1/6} \Lambda \delta^{1/3} \to  19 \alpha_{\theta f}^{4/3}~\mbox{TeV}~.
\end{equation}
Here we have taken $\gs^{\rm dec} \simeq 2\times10^2$, assuming that the preonic sector has a comparable number of degrees of freedom to the SM sector (cf. the toy model above). The decoupling temperature is close to the preonic confinement scale, and if $T_{\rm dec} < \Lambda$ -- a scenario of central interest below --  then $\gs^{\rm dec}$ could be slightly smaller. This, however, has only a marginal effect, so we keep $\gs^{\rm dec} \simeq 2\times10^2$ hereafter. Note also that the $2 \to 2$ decoupling of laurinos from the hidden pions, $T_{\rm dec,\Pi} \sim (\gs^{\rm dec})^{1/2} \Lambda \delta \to 39 \alpha_{\theta f}^{8/3}\alpha_m^{2/3}$~TeV, can be slightly smaller than $T_{\rm dec}$ if $\fdm$ is small enough. However, this does not significantly affect the thermal history.

For the decoupling temperature \eqref{eqn:TDC}, the laurinos decouple while relativistic. Consequently, they have a relic number density to entropy ratio
\begin{equation}
	\label{eqn:LRD}
	Y_{d} = Y^{\rm eq}_{\rm rel} \equiv 135\zeta(3)/2\pi^4\gs^{\rm dec}~,
\end{equation}
for each laurino Dirac species. For keV scale masses, this results in an overclosed universe, unless there is a significant source of entropy production, that dilutes the laurino abundance.

\subsection{Entropy Production}  
A long-lived heavy composite neutrino $N_{L,R}$ is a possible source of this entropy. However, the heavy neutrinos typically rapidly decay with $\Lambda$-scale widths to the chiral $n_R$ or hidden pions. They may also rapidly annihilate into the hidden pions via $N_RN^c_L \to 2\Pi$ or into the chiral bound states via $N_R^c N_R \to n_R^c n_R$ with $\Lambda$-scale scalar or vector mediators respectively.  These annihilation and decay channels ensure the confined sector comprises just $n_R$ and pions shortly after confinement. Let us consider whether these channels may be suppressed: the goal of this discussion is to develop a scenario in which the dominant decay of a heavy composite neutrino proceeds through ${N}_R \to {\nu_L} h$.

The mass term and right-handed vector coupling for the composite neutrinos can be written in the generic form $m_{Ji}N_L^{cJ}N_R^i$ and $v_{ij}{N_R^{c}}^i\gamma^\mu {N_R}^j$ respectively. Here lowercase (uppercase) Latin indices label the flavor of the right-handed (left-handed) composite neutrinos. By construction, for $J \in [1,n]$ we require $i \in [1,n+3]$ to ensure there are three chiral right-handed neutrinos.  For example, in the toy model, there are four flavors of $N_R^i \sim \chi_0^2\chi_2$ and one flavor of $N_L^{cJ} \sim \chi_2^3$.  For any $m_{Ji}$, there exists a unitary rotation of $N_L^J$ and $N_R^i$ such that three $N_R^i \sim n_R$ are explicitly massless, but generically the vector couplings $v_{ij}$ remain non-zero.

Alternatively, if one assumes $m_{Ji}$ and $v_{ij}$ are simultaneously democratic, i.e. $m_{Ji} = v_{ij} = 1$, then the unitary transformation to the mass basis also ensures $v_{i,j>n} = 0$. That is, there is no vector coupling to the chiral right-handed bound states: the vector annihilation channel, as well as the corresponding vector channel decay $N_R \to n_R n_R^c n_R$, vanishes. 

 The $\Lambda$-scale strong couplings are determined dominantly by the confining $\Gc$ structure. Democracy for $\Lambda$-scale composite neutrino interactions may then be a plausible assumption in the case that all the $N_R^i$ and $N_L^J$ are comprised of the same preonic irreducible representations respectively, as they are e.g. in the toy model. We therefore adopt this democracy assumption hereafter. In principle, this democracy should hold up to at most $\mathcal{O}(\epsilon^2)$-suppressed interactions, which decouple at $T_{\rm dec}$. The degree of suppression may be higher, however, depending on the UV theory above $M$. We assume in this discussion that these interactions can be neglected.

As we saw in the toy model above, it is possible for all the Dirac $N_{L,R}$ to moreover have negligible coupling to pions. This occurs if the preons that constitute the neutrinos are charged under a different gauge interaction to the preons that induce the hidden flavor symmetry breaking. For example, in the toy model the single heavy Dirac neutrino $ \{ \chi_0^2\chi_2, \chi_2^3\}$ has this property with respect to the vev  $\langle \chi_1\chi_3 \rangle$.

Let us now assume one of these heavy neutrinos, with democratically suppressed decays and negligible pion coupling, is also the lightest massive bound state in the spectrum, with mass $\gtrsim 3\Lambda$. We denote this heavy neutrino by $\Nl$. To a good approximation this flavor state is also a mass eigenstate after EWSB (cf. eq. \eqref{eqn:ME}). 

Being the lightest massive bound state, one sees from eq. \eqref{eqn:ME} that the dominant decay mode of the $\Nl$ neutrino is into the SM sector, via ${\Nl}_R \to {\nu^a_L}_i h$, $i = 1, 2, 3$. These processes are generated by diagrams like Fig.~\ref{fig:NLH}, with yukawa $\sim \epsilon^3/(4\pi)^2$.  The corresponding decay rate, including phase space factors, is
\begin{equation}
	\Gamma_{\Nl} \sim \frac{9 \tilde{m}\Lambda }{16 \pi}\frac{\epsilon^6}{(4\pi)^4} \to 6.2 \times 10^{-14}\frac{\alpha_{\Gamma}^2\tilde{m}}{(\alpha_m^7\alpha_{\theta f}^4)^{1/3}} ~\mbox{eV}~,
\end{equation}
where $\tilde{m} \equiv m_{\Nl}/3\Lambda$.  The corresponding lifetime is $\tau \lesssim 10^{-2}$~s. We see that $\Nl$ is long lived, and may therefore decay while out of equilibrium into the SM sector, potentially producing significant entropy dilution. 

Let us see if this is actually the case. First, it is essential to determine the $\Nl$ relic number density to entropy ratio, $Y_{\Nl}$. With reference to eqs. \eqref{eqn:LE}, \eqref{eqn:DD} and \eqref{eqn:TDC}, observe
\begin{equation}
	\label{eqn:LTD}
	\lambda \equiv 3\Lambda/T_{\rm dec} \sim 3\bigg(\!\frac{1}{\delta^{2}\gs^{\rm dec}}\!\bigg)^{1/6} \to 2.1\bigg(\frac{1}{\alpha _{\theta f}^2 \alpha _m}\bigg)^{1/3}.
\end{equation}
That is, $T_{\rm dec}$ and the confinement scale are close. The remarkable coincidence of these scales means that we can imagine $\Lambda \gtrsim T_{\rm dec}$. In this case, the heavy bound states can thermalize through $M$-scale interactions after or even during confinement -- the latter may occur if the CPT occurs slowly enough, in the fashion of QCD \cite{Aoki:2006we}. The $\Nl$ then freeze out of the plasma at $T_{\rm dec}$ with an equilibrium fiducial number density.

Since $m_{\Nl} \gtrsim 3\Lambda$, the $\Nl$ decouple from the thermal bath while non-relativistic. Normalizing with respect to the equilibrium relativistic abundance, the estimate of their abundance is correspondingly
\begin{equation}
	\label{eqn:YNH}
	Y_{\Nl} \simeq \frac{\sqrt{2\pi}}{3} Y^{\rm eq}_{\rm rel} (\lambda\tilde{m})^{3/2} e^{-\lambda\tilde{m}}~.
\end{equation}

The $\Nl$ come to dominate the energy density of the universe provided they decay late enough that $\rho_{\rm rad} \ll \rho_{\Nl}$. The putative matter-radiation equality occurs at temperature $T_{\rm eq} \sim m_{\Nl} Y_{\Nl}$. Above $T_{\rm eq}$, the universe is in a radiation dominated epoch, so that
\begin{equation}
	\frac{\Gamma_{\Nl}}{H}\bigg|_{T \gtrsim T_{\rm eq}} \lesssim \frac{\Gamma_{\Nl} \Mpl}{(m_{\Nl} Y_{\Nl})^2}  \to 3 \times 10^{-9}\frac{e^{2\lambda \tilde{m}}}{\tilde{m}^4}~. 
\end{equation}
Provided $m_{\Nl}$ is not too large compared to $T_{\rm dec}$, we see that the nonrelativistic $\Nl$ live long enough to dominate the energy density of the universe.

The $\Nl$ decay dominantly to the SM sector while far out of equilibrium. This leads to reheating of the SM sector up to a temperature $T_{\rm rh}$.  Energy conservation ensures that the reheated SM energy density $\rho_{\rm rh} \sim \Gamma_{\Nl}^2\Mpl^2$. Assuming the daughter neutrinos and higgs can rapidly thermalize, it follows that the reheat temperature
\begin{equation}
	\label{eqn:TRH}
	T_{\rm rh}  \sim \bigg(\frac{30}{g_*^{\rm rh} \pi^2}\bigg)^{1/4} \frac{\Lambda \epsilon }{\delta^{1/2}}  \to 8.7 \frac{\alpha_\Gamma\sqrt{\tilde{m}}}{(\alpha_m^7\alpha_{\theta f}^4)^{1/6}}~\mbox{MeV}~,
\end{equation}
with $g_*^{\rm rh} \simeq 10$. Note that both the neutrinos and higgs do thermalize rapidly at this temperature. Furthermore, the $\Nl$ decay reheats the SM sector above the BBN temperature, so that nucleosynthesis is not affected by this entropy production mechanism. $T_{\rm rh}$ is also well-below the non-resonant production temperature, so we can neglect this source of production even if $\fdm$ is small.

Finally, the entropy production estimate itself \cite{Kolb:1989}
\begin{align}
	\gamma &\equiv \frac{S_f}{S_i} \simeq 1.83 \frac{(g_*^{\rm rh})^{1/4} m_{\Nl} Y_{\Nl}}{\Gamma^{1/2}_{\Nl}\Mpl^{1/2}} \notag\\
	&\to 1.1 \times 10^7~Y_{\rm rel}^{\rm eq} \tilde{m}^2 e^{-\lambda \tilde{m}} (\alpha_m\alpha_{\theta f})^{1/3}\alpha_\Gamma^{-1}~. \label{eqn:EP}
\end{align}

\subsection{DM Properties} 
After entropy dilution, the relic density DM fraction for a single laurino species of mass $m_d$ becomes
\begin{equation}
	\label{eqn:FDM}
	\frac{\Omega_d}{\Omega_{\rm DM}} = \frac{m_d (Y_d/\gamma) s_0}{\rho_{\rm DM}}~.
\end{equation}	 
By construction $\fdm = \Omega_d/\Omega_{\rm DM}$. Applying eqs \eqref{eqn:LRD}, \eqref{eqn:LTD} and \eqref{eqn:EP} together with the observed values $s_0 \simeq 2.89\times10^3$ cm$^{-3}$, $\rho_c \simeq 10.5 h^{2}$cm$^{-3}$keV, and $\Omega_{\rm DM} \simeq 0.119h^{-2}$ \cite{pdg:2012}, one generates a transcendental equation for $\fdm$, viz.
\begin{multline}
	\label{eqn:FDME}
	\fdm  \simeq 1.5 \times 10^{-3}\frac{\alpha_\Gamma}{\tilde{m}^2\big[\fdm\alpha_m^2\alpha_{\theta}^2\big]^{1/6}}\\
	 \qquad \times \exp\bigg[2.1\tilde{m} \bigg(\frac{1}{\fdm\alpha _{\theta}^2 \alpha _m}\bigg)^{1/3}\bigg]~. 
\end{multline}

\begin{figure}[t]
\includegraphics[width=8cm]{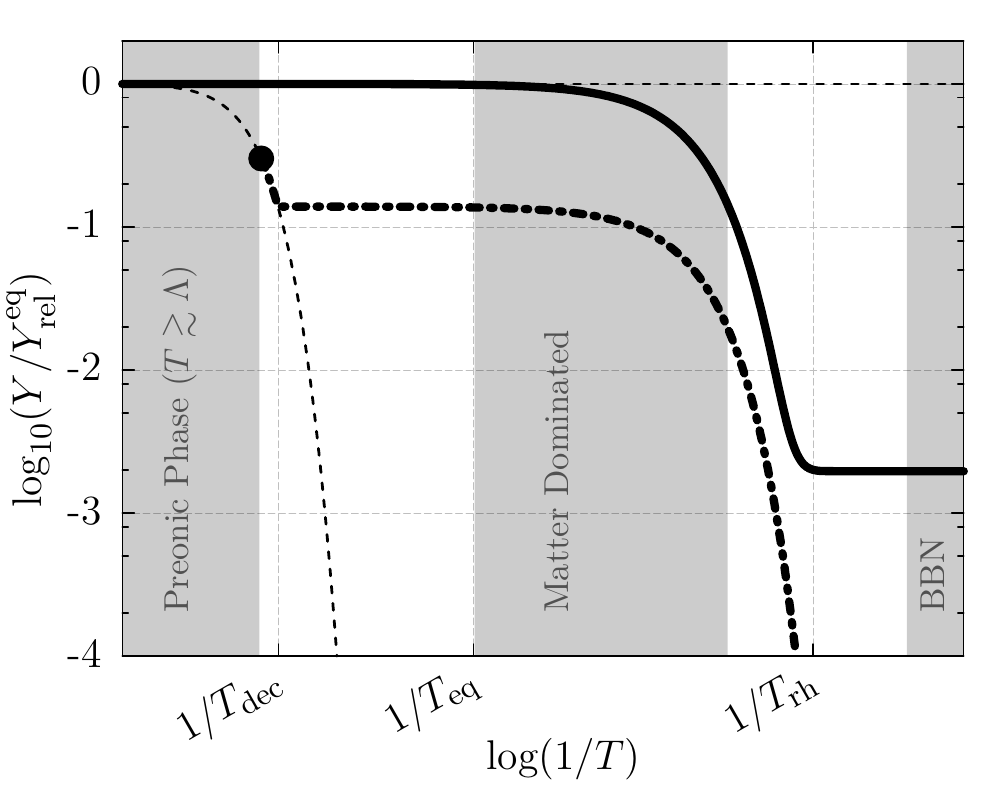}
\caption{Schematic thermal history of the laurino abundance (heavy black) and heavy composite neutrino abundance (heavy dot-dashed), compared to their equilibrium abundances (light dashed). Abundances are normalized by $Y^{\rm eq}_{\rm rel}$ \eqref{eqn:LRD}. Condensation and thermalization of $N^h$ occurs at $T \simeq \Lambda$ (black dot). This is followed by decoupling at $T_{\rm dec}$. Late decays of $N^h$ in the matter dominated non-equilibrium phase lead to entropy production and laurino dilution, with reheating of the SM plasma at $T_{\rm rh}$.}
\label{fig:STH}
\end{figure}

If all $\mathcal{O}(1)$ numbers are close to unity, then the solution of eq. \eqref{eqn:FDM} is 
\begin{equation}
	\label{eqn:FDMCV}
	\fdm \simeq 0.13~.
\end{equation}
This central value lifts $T_{\rm rh} \sim18$~MeV, sets the dilution factor $\gamma \sim 5 \times 10^2$, while lowering $T_{\rm dec} \sim 5$~TeV and $\Lambda \sim 7$~TeV. We see that $\Lambda \gtrsim T_{\rm dec}$, as required for this picture. A schematic summary of the thermal history of the laurinos and heavy composite neutrinos for these central values is shown in Fig. \ref{fig:STH}.

The exponential in eq. \eqref{eqn:FDME} ensures that mild choices for the $\mathcal{O}(1)$ parameters -- e.g. $\alpha_\theta \sim 1/3$ --  can easily lift $\fdm$, while simultaneously lifting the value of $T_{\rm rh}$, and lowering $T_{\rm dec}$ below $\Lambda$. Put in other words, we see that this framework is consistent with a single laurino species comprising an $\mathcal{O}(1)$ fraction, or perhaps all, of the observed DM. In the former case, the framework readily admits the possibility of other keV laurino species, whose x-ray lines are yet to be observed, or other, different sources of DM.  (It is amusing to note that the toy model predicts six laurino species.) 

One might be concerned that a small $\fdm$ may lift $\theta_f$ into the non-resonant production regime: nominally, for $\theta_d \simeq 4\times10^{-6}$, one requires $\fdm \gtrsim 0.4$ for non-resonant production to account for less than 5\% of the DM (see e.g. \cite{Kusenko:2009up}). However, we can neglect non-resonant production, because $T_{\rm rh}$ is well below the dominant non-resonant production temperature, $\sim 150(m_d/\mbox{keV})^{1/3}$~MeV \cite{Dodelson:1994sn}. For $\fdm \gtrsim 0.1$, the large entropy dilution \eqref{eqn:EP} ensures any non-resonantly produced relic is negligible.

After decoupling at $T_{\rm dec}$, the laurinos are cooled with respect to the active neutrino plasma by decoupling of SM species down to $T_{\rm rh}$, followed by entropy dilution by $\Nl$ decays. The present laurino temperature compared to the neutrino temperature, for the central values in eq. \eqref{eqn:EP}, treating $\gs^{\rm dec}\gamma$ as equivalent to the relativistic degrees of freedom at laurino decoupling,
\begin{equation}
	T^0_d  \sim \bigg[\frac{\gs^{\rm dec,\nu}}{\gs^{\rm dec}\gamma }\bigg]^{1/3}T^0_\nu \to 0.04\fdm^{1/3}~T_\nu^0~.
\end{equation}
Here we have applied eqs. \eqref{eqn:LRD} and \eqref{eqn:FDM}, from which one sees that the temperature scales as $T^0_d \sim \fdm^{1/3}$. The laurinos are therefore a little cooler compared to other sterile neutrino WDM candidates, and may evade current structure formation bounds. In particular, the Lyman-$\alpha$ lower bound conservatively requires a 100\% WDM thermal mass $m_{\rm th} > 3.3$~keV \cite{Viel:2013fqw,Horiuchi:2013noa}, equivalent to a Dodelson-Widrow mass of $22$~keV. For $\fdm <  1$, and noting the laurinos have four internal degrees of freedom, the corresponding effective thermal mass bound is $m_d >  2.8 \fdm^{1/4}$~keV, which is well below the laurino mass. Similarly, current N-body simulation bounds require a 100\% WDM thermal mass $m_{\rm th} > 2.6$~keV \cite{Schultz:2014eia}, which becomes $m_d > 2.2\fdm^{1/4}$~keV.

The corresponding free-streaming or Jeans length for the laurinos can be explicitly extracted from standard WDM results. One finds then that at the present epoch \cite{Kolb:1989,Markovic:2013iza},
\begin{equation}
	\lambda_{\rm fs} \sim 1.3\bigg[ \frac{T_d^0}{T_\nu^0}\bigg]\bigg[\frac{ \mbox{keV}}{m_d}\bigg] \mbox{Mpc} \to 7.4\fdm^{1/3}~\mbox{kpc}~.
\end{equation}
Equivalently, the free-streaming mass for the laurinos is 
\begin{equation}
	M_{\rm fs} \equiv  \pi \rho_{\rm m} \lambda_{\rm fs}^3/6 \to 1.9 \times 10^4\fdm M_{\odot}~.
\end{equation}

The right-handed sterile neutrinos $\nu^a_R \sim n_R$ and hidden pions have a similarly diluted temperature, since they decouple at the same time as the laurinos, i.e. at $T_{\rm dec}$, and are not reheated by $\Nl$ decays. They may, however, be reheated during the CPT, by at most an $\mathcal{O}(1)$ temperature ratio. Nevertheless, still $T_{\nu_R, \Pi, \xi} \ll T_\nu$ at the BBN and CMB epochs, which follow reheating. As a consequence, the contribution of the right-handed neutrinos, pions and laurinos to the effective neutrino degrees of freedom at these epochs is negligible.

Finally, let us briefly consider whether this framework admits Dirac-type leptogenesis, given its thermal history. Although net lepton number is conserved, a lepton asymmetry may be generated by a process that separates lepton number into decoupled sectors. For example, an electroweak lepton asymmetry may be produced by a CP-violating decay of a heavy composite scalar, $\Theta \to \xi^c_R\nu_L$, that occurs after SM-laurino decoupling. This decay proceeds though an $M$-scale interaction, with rate $\Gamma_\Theta \sim \tilde{m}_{\Theta}^5\Lambda \epsilon^4/(16\pi^2)^2$. In the absence of other decay channels, the $\Theta$ lifetime $\sim 5 \times10^{-13}$~s for $\tilde{m}_{\Theta} \equiv m_{\Theta}/\Lambda \sim 5$, say.  This is just long enough for $\Theta$ to decay after $T_{\rm dec}$ but before EWSB, raising the further possibility that the resulting electroweak lepton asymmetry could be rotated into the baryons through electroweak sphalerons. Estimating this baryon asymmetry is beyond the scope of this work, but note that it will be subsequently diluted by the entropy production.

Despite this mildly attractive narrative for leptogenesis, it is difficult to conceive of composite scalars in this framework that do not also couple and decay to the hidden pions through $\Lambda$-scale interactions; certainly the toy model does not contain such scalars. We therefore leave the construction of leptogenetic models to the future.

\section{Conclusions} 
The observation that the recently detected $3.5$~keV x-ray line exhibits $m_d\theta_d \sim \mbox{few}\times10^{-2}$~eV -- a plausible scale for the active neutrinos -- is strikingly suggestive that the x-ray source is a sterile Dirac neutrino that mixes with the actives. In this Note we have reviewed a framework that dynamically produces both light active and keV sterile Dirac neutrinos, with the appropriate mixing angles to be the x-ray source. The central idea is that the right-handed active neutrinos are composite degrees of freedom, while the elementary keV neutrinos obtain their mass through irrelevant couplings to the chiral symmetry breaking condensate, in the fashion of extended Technicolor.

With special choices for the hidden flavor structure, and assuming democratic couplings for the composite neutrino degrees of freedom, this framework may automatically include long-lived heavy composite neutrinos. These heavy neutrinos decay into the SM sector after dominating the energy density of the universe, at an epoch such that: the reheating produced by the decay reheats the SM sector above the BBN temperature; thermal relic keV neutrinos can be diluted down to an $\mathcal{O}(1)$ fraction of the observed DM abundance; the diluted relic temperature is somewhat lower than the usual sterile neutrino WDM candidates, and satisfies current structure formation bounds. We believe this framework is therefore an intriguing possibility to account for the observed DM.

A toy model that exhibits the central features of this model was also presented, though we expect there may be other possibilities, which could be the subject of future investigation. 

\acknowledgments
The authors thank Simone Alioli, Tom Banks, Francesco D'Eramo, Howie Haber, Zoltan Ligeti, Joel Primack, Joshua Ruderman and Jon Walsh for helpful discussions. We particularly thank Duccio Pappadopulo for providing valuable comments during the preparation of this work. The work of DR is supported by the U.S. National Science Foundation under Grant No.~PHY-1002399. The work of YT is supported by the Department of Energy under Grant DE-FG02-91ER40674. The work of YT was also supported by the hospitality of the Center for Future High Energy Physics in Beijing, China.


%

\end{document}